%% file: root.tex
\documentclass[letterpaper, 10 pt, conference]{ieeeconf}  
\IEEEoverridecommandlockouts                              

\title{\LARGE \bf
Strategically Robust Game-Theoretic Multi-Agent Trajectory Optimization
}

\author{Victor L. Qin$^{1*}$, Nicolas Lanzetti$^{2}$, Saverio Bolognani$^{3}$, and Hamsa Balakrishnan$^{1}$ 
\thanks{%
V. Qin would like to thank Geoffrey Ding for helpful conversations. V. Qin is supported by the National Science Foundation Graduate Research Fellowship Program under Grant~\#2141064, by NCCR Automation, grant agreement 51NF40\_225155 from the Swiss National Science Foundation, and in part by NASA Grant~\#80NSSC23M0220. N. Lanzetti is supported by the PIMCO Fellows Program, by the National Science Foundation under Grant~\#CCF-2326609, and by the Resnick
Sustainability Institute.
Any opinions, findings, and conclusions or recommendations expressed in this material are those of the authors and do not necessarily reflect the views of the National Science Foundation or of the National Aeronautics and Space Administration.}
\thanks{$^{1}$Department of Aeronautics and Astronautics, MIT.}%
\thanks{$^{2}$Department of Computing and Mathematical Sciences, Caltech.}%
\thanks{$^{3}$Automatic Control Laboratory, ETH Zurich.}
\thanks{$^{*}$Corresponding author, {\tt\small victorqi@mit.edu}}
}

\input{packages}

\begin{document}

\maketitle
\thispagestyle{empty}
\pagestyle{empty}

\begin{abstract}
Aviation authorities worldwide expect Advanced Air Mobility (AAM) traffic management to be decentralized among service providers, requiring AAM flights to autonomously plan trajectories by predicting other flights' control inputs rather than relying on centralized coordination. Game-theoretic approaches that formulate multi-agent collision avoidance as an exact dynamic potential game can efficiently find open-loop equilibria, but they assume that agents exactly follow their equilibrium trajectories---an unrealistic assumption given uncertainties in actuation, perception, and computation. We propose a strategically robust formulation where each agent protects against a fictitious adversary that, for each timestep, perturbs other agents' control inputs within a bounded budget to minimize distance at that timestep. We show that, under reasonable assumptions on agents' distance cost and robustness levels, the strategically robust game remains an exact dynamic potential game and admits a quasi-closed-form solution to the inner adversarial problem for linear dynamics, which limits computational overhead.
Experiments with up to eight agents using logarithmic distance costs show that strategic robustness selects more robust trajectories in high-collision-risk configurations while leaving low-risk trajectories nearly unchanged, with only a modest increase in runtime.
\end{abstract}

\input{sections/1_introduction}

\input{sections/2_model}

\input{sections/3_solving}
\input{sections/4_results}
\input{sections/5_conclusion}

{\footnotesize      
\bibliographystyle{IEEEconf}
\bibliography{ref}
}

\appendix
\input{sections/a_proof}

\input{sections/a_derivation}


\end{document}

%% file: packages.tex
\usepackage{graphicx} 
\usepackage{multirow}
\usepackage{bbm}
\usepackage{caption}
\usepackage{subcaption}
\usepackage{booktabs}
\usepackage{dblfloatfix}   

\usepackage{amsthm}
\usepackage{amssymb}
\usepackage{mathtools}
\usepackage[thinc]{esdiff}
\usepackage{esvect}
\usepackage{tikz}
\usepackage{fancyhdr}
\usepackage{bm}
\usepackage{algorithm,algpseudocode}
\usepackage{csquotes}
\usepackage{xcolor}
\usepackage{soul}
\usepackage{derivative}
\usepackage{microtype}
\usepackage{xurl}
\AtBeginDocument{%
  \abovedisplayskip=5pt plus 2pt minus 2pt
  \belowdisplayskip=5pt plus 2pt minus 2pt
  \abovedisplayshortskip=2pt plus 1pt
  \belowdisplayshortskip=3pt plus 1pt minus 1pt
}

\makeatletter
\let\NAT@parse\undefined
\makeatother
\usepackage[numbers,sort]{natbib}
\usepackage[hidelinks]{hyperref}
\usepackage{doi}
\usepackage[capitalize]{cleveref}
\crefname{assume}{Assumption}{Assumptions}
\Crefname{assume}{Assumption}{Assumptions}
\crefname{prop}{Proposition}{Propositions}
\Crefname{prop}{Proposition}{Propositions}

\newcommand{\R}{\mathbb{R}}
\newcommand{\Nc}{\mathcal{N}}

\newcommand{\trans}[1]{{#1}^\top}

\newcommand{\sta}[1]{{#1}^\star}

\newtheorem{thm}{Theorem}

\newtheorem{lem}{Lemma}
\newtheorem{defn}{Definition}

\newtheorem{assume}{Assumption}

\theoremstyle{remark}

%% file: sections/1_introduction.tex
\section{Introduction}\label{sec:intro}

Advanced Air Mobility (AAM) vehicles, such as unmanned aircraft systems (UASs) and electric vertical take-off and landing aircraft (eVTOLs), could transform urban and regional transportation~\cite{mckinsey_and_company_perspectives_2022,greenawalt2026primeair}. However, AAM operations pose a challenge for current air traffic management systems, as they are expected to operate autonomously on demand between many urban destinations. National aviation authorities worldwide envision AAM traffic management services being provided by several service providers in a region~\cite{faaUtmImplementationPlan,faaUamConops}.

With the decentralization of traffic management responsibilities, autonomous AAM flights cannot rely on a centralized authority to guarantee collision avoidance.
Instead, they must solve a game-theoretic trajectory optimization problem, where each flight (each agent) finds the most efficient trajectory while avoiding collisions with other agents. Standard game-theoretic methods that solve for the Nash equilibrium assume that each agent strictly adheres to its Nash equilibrium trajectory---a difficult assumption given the uncertainties inherent in aviation, even if agents are not adversarial (e.g., perception at night, miscalibrated sensors or systems, pilot errors)---and provide no guarantees when agents deviate.
Clearly, misspecified control inputs from other AAM flights, due to factors like limited computation or partial information, could lead to catastrophic collisions.

It is thus natural for flights to seek protection against such deviations, i.e., against uncertainty in the control inputs of other flights.
Such strategic uncertainty---in contrast to the standard uncertainty in robust control or optimization---is not \emph{exogenous} (e.g., against environmental noise as in robust control) but \emph{endogenous} to the flights, where the control inputs of a flight affect those of the other flights against which they seek protection. Placing the uncertainty on control inputs rather than directly on states also guarantees that every deviation that is being protected against is dynamically feasible for the deviating flight.

\begin{figure}[!t]
    \centering
    \includegraphics[width=.9\columnwidth]{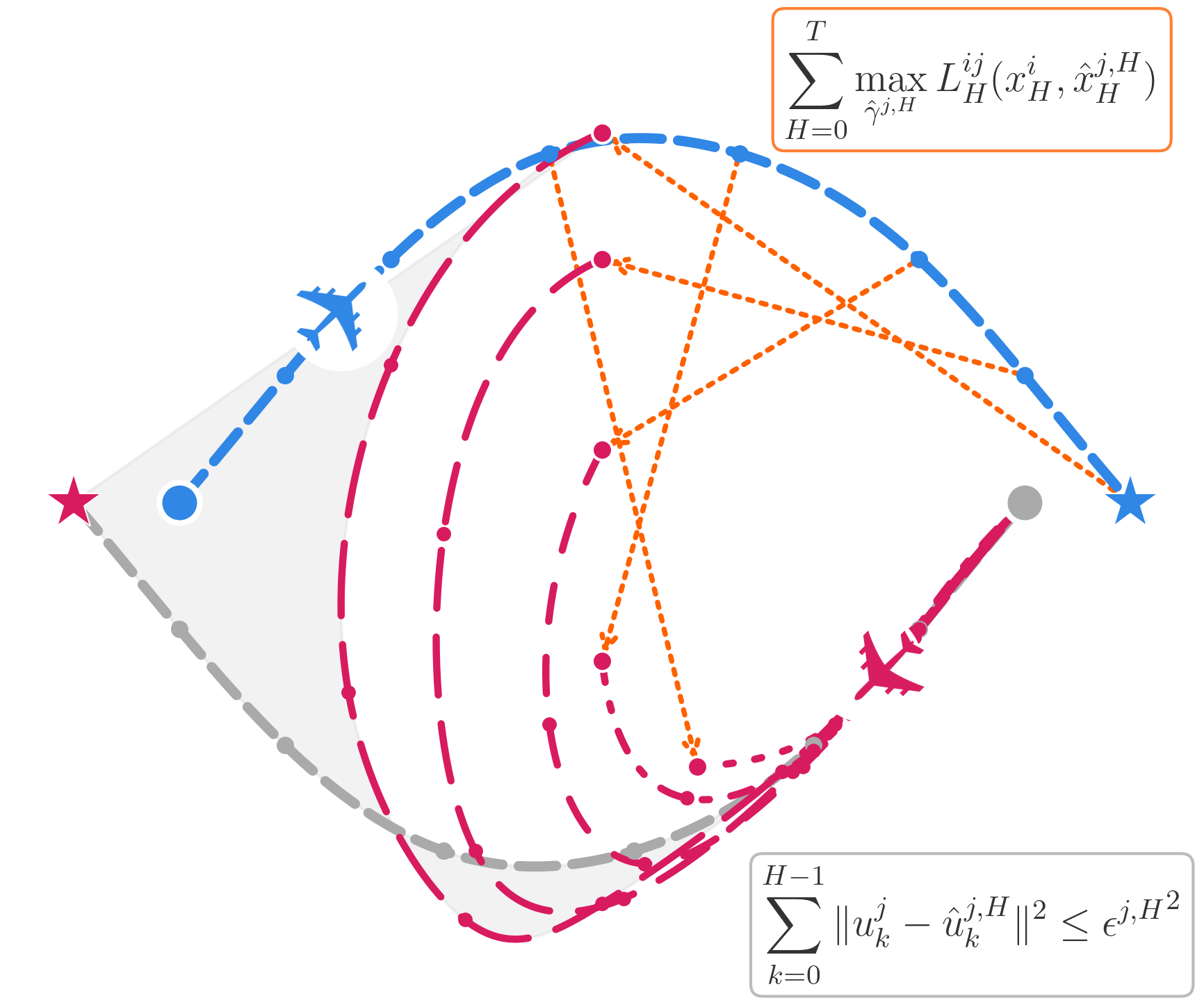}
    \caption{
   Illustration of the strategically robust approach for a two-agent scenario. The ego agent (blue trajectory) plans against a fictitious adversary that, for each timestep $1 \leq H \leq T$, perturbs the other agent's control inputs away from its nominal trajectory (gray) to minimize the distance between agents at that timestep (orange). The adversary's worst-case trajectory for each timestep (red) is subject to a budget constraint on the total control input deviation ($\sum_{k=0}^{H-1} \Vert u^{j}_k - \hat{u}^{j,H}_k \Vert^2 \leq {\epsilon^{j,H}}^2$, gray shaded region). See~\eqref{eq:adv_max_defn}.}
   \label{fig:cover}
   \vspace{-0.5em}
\end{figure}

To achieve this goal, we adopt a strategically robust approach, introduced for static games~\cite{lanzetti_nicholas_strategically_2025}, and propose that each agent minimizes its control cost against a fictitious adversary.
The key idea is illustrated in~\cref{fig:cover}. Each agent plans its own trajectory while anticipating that other agents may not exactly follow their nominal trajectories, by assuming that a fictitious adversary, at each step, perturbs the nominal control inputs of each of the other agents so as to minimize the distances to the ego agent---that is, to create the worst-case collision scenario. Crucially, the adversary is budget-constrained: for each of the other agents, the total deviation from its nominal control inputs cannot exceed a prescribed budget. By optimizing against this worst-case scenario at every timestep, each agent obtains a trajectory that is robust to bounded perturbations in the other agents' control inputs.

Agents can tune their desired level of robustness through this budget. If the budget is reduced to zero, the fictitious adversary is forced to replicate the control inputs of the other flights, recovering the standard Nash equilibrium in dynamic games.
As the budget is increased, players assign more power to their fictitious adversary, thereby increasing robustness to perturbations in the strategies of the other players. The budget can therefore be directly interpreted as the level of robustness.

While attractive in spirit, integrating strategic robustness into multi-agent trajectory optimization poses significant challenges. Even in the absence of strategic robustness, finding equilibria in dynamic games involves solving coupled nonlinear optimization problems and is computationally expensive.
Recent work~\cite{bhatt_efficient_2023, bhatt2025strategic} identified that, under mild assumptions, multi-agent trajectory optimization can be structured as a dynamic potential game~\cite{monderer_potential_1996, zazo_dynamic_2016, anardi_urbandriving_2021}, which effectively allows us to find multi-agent equilibrium trajectories by solving a single constrained optimization problem.
Whether strategic robustness preserves this structure without altering its favorable computational properties is the main challenge in deploying it for multi-agent trajectory optimization.

In this paper, we answer this question affirmatively. In particular, our contribution is threefold:
\begin{enumerate}
    \item We show that, under natural assumptions on the distance cost, using strategic robustness in exact dynamic potential games leads to another dynamic potential game with an appropriately modified potential function.
    \item While the modified potential function requires solving a worst-case optimization problem, we obtain a quasi-closed-form solution to this adversarial maximization that can be quickly computed online.
    \item Through various numerical examples, we show that our strategically robust method does not significantly add to the runtime of the trajectory optimization.
\end{enumerate}

\subsection{Related work}

\paragraph*{Game-theoretic planners}
ALGAMES \cite{cleach_algames_2020} solves for generalized Nash equilibria using Newton's method on KKT conditions, while dynamic potential games \cite{zazo_dynamic_2016} enable faster convergence by reducing the game to a single optimization problem \cite{bhatt_efficient_2023, bhatt2025strategic, kavuncu2021potential, williams_distributed_2023}. This optimization problem is solved online using iterative trajectory optimization methods based on linear-quadratic approximations, such as iLQR \cite{li_iterative_2004} and ALTRO \cite{howell_altro_2019}. While dynamic potential games are more restrictive than generalized Nash equilibrium games, multi-agent trajectory problems naturally fit in the dynamic potential game structure~\cite{zazo_dynamic_2016, bhatt_efficient_2023}. However, these planners assume perfect knowledge of other agents' cost functions and dynamics.

\paragraph*{Exogenous uncertainty}
In decision theory, there are various approaches that are robust against exogenous uncertainty, including distributionally robust optimization~\cite{kuhn2025distributionally} and risk measures~\cite{follmer_convex_2002}.
Risk measures have been applied to risk-aware robotics \cite{akella2025risk, ryu2024integrating}.
Risk-sensitive iLQR games \cite{wang2020game} incorporate noise in dynamics via entropic risk. Yet these approaches are primarily robust to exogenous uncertainty, and not to strategic uncertainty which is instead endogenous.

\paragraph*{Strategically robust game theory}
To protect against strategic uncertainty (i.e., uncertainty about the other players), we adopt the strategically robust game-theoretic approach \cite{lanzetti_nicholas_strategically_2025}.
In strategically robust game theory, agents make decisions against a fictitious agent that aims to inflict maximum damage but is constrained to lie within a prescribed distance of the mixed strategy of all other players.
This way, strategically robust equilibria interpolate between Nash and security equilibria.
The open-loop strategically robust equilibria we use for this work are precisely inspired by this philosophy and can be interpreted as pure strategically robust equilibria in open-loop dynamic games.
Under natural assumptions on the interagent cost structure, we show that our strategically robust game retains the exact dynamic potential game property \cite{zazo_dynamic_2016}, thereby preserving the computational advantages of solving a single optimization problem while adding robustness.
More broadly, our work subscribes to a growing body of recent literature that leverages strategic robustness and risk aversion, sometimes combined with bounded rationality, in multi-agent settings to improve robustness, tractability, and sometimes even collaboration~\cite{mazumdar2025tractable,zhang2025convergent,qu2026training,feik2026strategically,velasevic2026strategically}.

%% file: sections/2_model.tex
\section{Strategically Robust Trajectory Optimization}\label{sec:model}

We consider a multi-agent trajectory optimization problem, where agents $i \in \{1, 2, \dots, N \} \coloneqq \Nc$ optimize their trajectories over a time horizon $0 \leq k \leq T$. Let $x^i_k \in \R^n$ and $u^i_k \in \R^m$ be the state and control input of agent $i$ at time $k$; $x_k$ and $u_k$ are the concatenations of the state and input of all agents at time $k$; $x^i$ and $u^i$ are the concatenations of the state and input of agent $i$ for all timesteps; and $x = \{ x_k \}_{0 \leq k \leq T }$ and $u = \{ u_k \}_{0\leq k \leq T-1}$.
The agent dynamics are linear and identical across all agents:
\begin{assume}[Dynamics] \label{assume:dynamics}
    The agent dynamics are
    \begin{equation*}
        x^i_{k+1} = A x^i_k + B u^i_k \quad \forall i \in \Nc.
    \end{equation*}
\end{assume}
We define the set of feasible trajectories for an agent given some fixed initial state $x^i_0$ to be $C^i$, and the set of trajectories for all agents to be $C\coloneqq\prod_{i \in \Nc} C^i$:
\begin{equation*}
\begin{aligned}
    C^i =& \{ (x^i_0, \dots, x^i_T, u^i_0, \dots, u^i_{T-1}) | \\
    &x^i_{k+1} = A x^i_k + B u^i_k \;\; \forall  k \in [0, T-1] \}.
\end{aligned}
\end{equation*}
Given other agents' control inputs $u^{-i}$ and fixed initial state $x^i_0$, each agent seeks to minimize the control cost
\begin{equation}\label{eq:nominal_cost_function}
\begin{aligned}
    \min_{u^i}\;& J^i(x_0, u) \coloneqq L^i_T(x_T) + \sum_{k = 0}^{T-1} L^i_k(x_k, u_k) \\
    \text{s.t.} \;\; & x^l_{k+1} = A x^l_k + B u^l_k
    \quad \forall l \in \Nc, \; k \in [0, T-1] ,
\end{aligned}
\end{equation}
where we assume all functions are continuously differentiable.
As in~\cite{bhatt2025strategic}, we decompose $L^i_k(x_k, u_k)$ into a private component $L^{ii}_k$, which depends on the agent state and input, and an interagent component $L^{ij}_k$ for $j \in \Nc \setminus \{i\}$, which captures effects such as collision avoidance and depends on the states $x_k^i$ and $x_k^j$ of agents $i$ and $j$:
\begin{equation}
\begin{aligned}\label{eq:cost_decomposition}
    L^i_k(x_k,u_k) &= L^{ii}_k(x^i_k, u^i_k) + \sum_{j \in \Nc \setminus \{i\}} L^{ij}_k(x^i_k, x^j_k) \\
    L^i_T(x_T) &= L^{ii}_T(x^i_T) + \sum_{j \in \Nc \setminus \{i\}} L^{ij}_T(x^i_T, x^j_T).
\end{aligned}
\end{equation}
For example, the private cost can be a standard quadratic cost on the distance to a goal $x^i_f$, i.e., $L^{ii}_k(x_k^i, u^i_k) = \trans{(x^i_k - x^i_f)} Q^i_k (x^i_k - x^i_f) + \trans{u^i_k} R^i_k u^i_k$, while the interagent cost $L^{ij}_k(x_k^i,x_k^j) = -c \ln(\Vert x^i_k - x^j_k\Vert^2 + \eta)$ pushes agents apart to avoid collisions, given parameters $c > 0, 1 \gg \eta > 0$.\footnote{The regularizer $\eta$ simply keeps $L^{ij}_k(x^i_k, x^j_k)$ finite so that the cost remains well-defined.}

\subsection{Game-theoretic solution concept}

We consider a finite horizon open-loop trajectory optimization problem with horizon $T$, where agent $i$ decides on a strategy $\gamma^i(x_0) = \{ u^i_k \}_{k\in [0, T-1]} \in \Gamma^i$, a sequence of control inputs that generates a trajectory $x^i = \{ x^i_k \}_{k \in [0, T]}$ such that $(x^i, \gamma^i(x_0) ) \in C^i$, with the initial state $x^i_0$ given by $x_0$.  The joint strategy of all players is $\gamma(x_0) =  (\gamma^1(x_0), \dots, \gamma^N(x_0))$, where $\Gamma=\prod_{i \in \Nc} \Gamma^i$ is the space of all joint strategies; we will write $\gamma^i(x_0, k) = u^i_k$.
We write the cost to an agent as $J^i(x_0, \{ \gamma^i(x_0), \gamma^{-i}(x_0) \})$, where the trajectory $x$ is generated from $x_0$ by the strategies in the second argument.
\begin{defn}[Nash equilibrium]
    An open-loop Nash equilibrium for a game $G = (\Nc, \{J^i\}_{i \in \Nc}, \Gamma) $ given initial conditions $x_0$ is a set of strategies $\sta{\gamma} = (\sta{\gamma^1}, \dots, \sta{\gamma^N}) \in \Gamma$ with $(x,\sta{\gamma}(x_0)) \in C$ such that for all agents $i \in \Nc, \gamma^i \in \Gamma^i$:
    \begin{equation*}
        J^i(x_0, \{ \sta{\gamma^i}(x_0), \sta{\gamma^{-i}}(x_0) \})
        \leq J^i(x_0, \{ \gamma^i(x_0), \sta{\gamma^{-i}}(x_0) \}).
    \end{equation*}
\end{defn}

The computation of open-loop Nash equilibria involves solving nonlinear equations coupled between players and is therefore computationally challenging.
However, if the interagent costs satisfy symmetry properties---i.e., $L^{ij}_k(x^i_k,x^j_k)=L^{ji}_k(x^j_k,x^i_k)$---then the game admits a potential function; see \cite{zazo_dynamic_2016,bhatt2025strategic,bhatt_efficient_2023} and the proof of our \cref{thm:sre_exact_PG} below for details.
\begin{defn}[Dynamic potential game]\label{defn:potential_game}
    A game $G=(\Nc, \{J^i\}_{i \in \Nc}, \Gamma)$ is an exact dynamic potential game if there exists a potential function $\Phi(x_0, \gamma(x_0))$ such that
    \begin{equation}
    \begin{aligned}
        &J^i(x_0, \{ \gamma^i(x_0), \gamma^{-i}(x_0) \}) \! - \! J^i(x_0, \{ \gamma^{i'}(x_0), \gamma^{-i}(x_0) \}) \\
         =&  \Phi(x_0, \{ \gamma^i(x_0), \gamma^{-i}(x_0) \}) \! - \! \Phi(x_0 , \{ \gamma^{i'}(x_0), \gamma^{-i}(x_0) \}).
    \end{aligned}
    \end{equation}
\end{defn}
If a game is an exact dynamic potential game, every minimizer of the potential function is an open-loop Nash equilibrium. Thus, whenever the potential attains its minimum (e.g. when it is coercive), an equilibrium exists and can be found by solving a single optimization problem~\cite{bhatt_efficient_2023, bhatt2025strategic}.
Dynamic potential games can also be defined as games where we can represent $J^i$ as the sum of a potential function and a term that is independent of the agent's own strategy; see [\citealp[Lemma 3]{zazo_dynamic_2016}] and [\citealp[Theorem 2.1]{voorneveld_congestion_1999}] for more details.
\begin{lem}[adapted from \protect{\cite[Theorem 2.1]{voorneveld_congestion_1999}}]\label{lem:coord_dummy}
    The game $G=(\Nc, \{J^i\}_{i \in \Nc}, \Gamma)$ is an exact dynamic potential game with potential function $\Phi$ if and only if there exist functions $\Theta^i$, $i \in \Nc$, such that
    \begin{equation}\label{eq:coord_dummy_lem}
        J^i(x_0, \gamma(x_0)) = \Phi(x_0, \gamma(x_0)) + \Theta^i(x_0, \gamma^{-i}(x_0))
    \end{equation}
    for all $i \in \Nc$ and all $\gamma \in \Gamma$.
\end{lem}

\subsection{Strategically robust equilibria}

The Nash equilibrium is a natural solution concept for such a multi-agent decision problem, but in practice agents might fear misbehavior from others, i.e., deviations in their control inputs.
We seek protection against such deviations by modifying the distance cost between ego agent $i$ and agent $j$ at time $H \in [1,T]$ as follows:\footnote{At $H=0$, $\hat{x}^{j,0}_0 = x^j_0$, so $\tilde{L}^{ij}_0 = L^{ij}_0$ and we recover the nominal cost.}
\begin{equation}\label{eq:adv_max_defn}
\begin{aligned}
    \tilde{L}^{ij}_H(x^i_H,& x^j_H) \\
   = \max_{\hat{\gamma}^{j,H}} \quad & L^{ij}_H(x^i_H, \hat{x}^{j,H}_H) \\
     \text{s.t.}
     \quad
     & \hat{x}^{j,H}_{k+1} = A \hat{x}^{j, H}_k + B \hat{u}^{j, H}_k \; \forall k \in[0, H-1] \\
     & \hat{x}^{j, H}_0 = x^j_0
     \\
    & \sum_{k=0}^{H-1} \Vert u^j_k - \hat{u}^{j, H}_k \Vert^2 \leq {\epsilon^{j,H}}^2.
\end{aligned}
\end{equation}
Each agent $i \in \Nc$ evaluates the distance cost against a fictitious adversary. This adversary selects, for each agent $j \in \Nc \setminus \{i\}$ and timestep $H \in [1, T]$, an open-loop policy $\hat{\gamma}^{j,H}$ with the goal of maximizing the distance cost for agent $i$ at each timestep $H$ (equivalently, minimizing the distance between agents $i$ and $j$ at time $H$), but cannot deviate in total by more than $\epsilon^{j,H}$ from the nominal control inputs $u^j_k, k\in [0, H-1]$.\footnote{
The budget constrains deviations \emph{from} $\{u^j_k\}_{k=0}^{H-1}$, so the adversary's reachable set at time $H$ is fixed, convex, and centered at $x^j_H$. Equivalently, under \cref{assume:monotonicity}, \eqref{eq:adv_monotonic_defn} below depends on agents $i$ and $j$ only through the relative position $z_H = x^i_H - x^j_H$. Hence $\tilde{L}^{ij}_H$ is well defined as written and continuously differentiable (since squared distance to a convex set is).}
As described in \cref{sec:intro}, $\epsilon^{j,H}$ is the robustness level: at $\epsilon^{j,H}=0$ the adversary is constrained to the nominal inputs $u^j_k$ and we recover the standard Nash equilibrium, while larger values protect against larger perturbations.

Given this robust distance cost, the strategically robust control cost of each agent $i$ is
\begin{equation}\label{eq:robust_cost}
\begin{aligned}
    &\tilde{J}^i(x_0, \{ \gamma^i(x_0), \gamma^{-i}(x_0) \}) \\
    =& \, L^{ii}_T (x^i_T) + \sum_{k = 0}^{T-1} L^{ii}_k(x^i_k, u^i_k) + \!\!\!\! \sum_{j \in \Nc \setminus \{i\}} \sum_{H = 0}^{T}  \tilde{L}^{ij}_H(x^i_H, x^j_H).
\end{aligned}
\end{equation}
With this, we can define strategically robust equilibria in our context as follows:

\begin{defn}[Open-loop strategically robust equilibrium]
    An open-loop strategically robust equilibrium for a game $\tilde{G} =(\Nc, \{\tilde{J}^i\}_{i \in \Nc}, \Gamma) $ given initial conditions $x_0$ is a set of strategies $\sta{\gamma} = (\sta{\gamma^1}, \dots, \sta{\gamma^N}) \in \Gamma$ with $(x, \sta{\gamma}(x_0)) \in C$ such that for all agents $i \in \Nc, \gamma^i \in \Gamma^i$ we have
    \begin{equation*}
        \tilde{J}^i(x_0, \{ \sta{\gamma^i}(x_0), \sta{\gamma^{-i}}(x_0) \})
        \leq
        \tilde{J}^i(x_0, \{ \gamma^i(x_0), \sta{\gamma^{-i}}(x_0) \}).
    \end{equation*}
\end{defn}

Note that our fictitious adversary is solving a target intercept problem with a separate adversarial trajectory $\hat{\gamma}^{j,H}$ given deviation budget $\epsilon^{j,H}$ for each timestep $H$.
This is deliberate: a collision between two aircraft at any one timestep is catastrophic.
Our fictitious adversary is at least as powerful as one committing to a single trajectory maximizing the more standard formulation of a summed distance cost over the entire horizon $[0, T]$, given the same budget.

%% file: sections/3_solving.tex
\section{Computation of Strategically Robust Equilibria}

We now show that the strategically robust game preserves the potential structure of the nominal game, which makes its equilibria efficiently computable.

\subsection{Strategic robustness in exact dynamic potential games}\label{ssec:sre_as_pg}

We make two further assumptions. First, we assume the interagent cost is a monotone function of the distance between agents.\footnote{Only the collision-relevant components of the state enter the distance cost; formally, $\Vert D(x^i_H - x^j_H) \Vert^2$ for a fixed selection matrix $D$. We take $D = I$ to lighten notation.}
\begin{assume}[Distance cost] \label{assume:monotonicity}
    The distance cost is $L^{ij}_k(x^i_k, x^j_k) = -\mu(\Vert x^i_k - x^j_k \Vert^2)$, where $\mu: \R_{\geq 0} \to \R$ is a monotonically increasing function.
\end{assume}

This assumption is natural for collision avoidance, where the cost should grow as agents approach one another.
The logarithmic distance penalty from~\cref{sec:model},
$L^{ij}_k(x^i_k, x^j_k) = -c \ln(\Vert x^i_k - x^j_k\Vert^2 + \eta)$,
satisfies this assumption.

Second, we assume players have the same robustness level.

\begin{assume}[Symmetric robustness] \label{assume:deviation}
    For all times $H \in [1, T]$, agents have the same robustness level, i.e.,
    $\epsilon^{i,H} = \epsilon^{j,H}$ for all $i, j \in \Nc$.
\end{assume}

Our main theoretical result follows.

\begin{thm}\label{thm:sre_exact_PG}
    Let $G=(\Nc, \{J^i\}_{i \in \Nc}, \Gamma)$ be a dynamic game with $J^i$ given by \eqref{eq:nominal_cost_function} and \eqref{eq:cost_decomposition}, and let \cref{assume:dynamics,assume:monotonicity} hold so that $G$ is an exact dynamic potential game.
    Suppose \cref{assume:deviation} holds.
    Then the strategically robust game $\tilde{G} = (\Nc, \{\tilde{J}^i\}_{i \in \Nc}, \Gamma) $ is also an exact dynamic potential game with potential function
    \begin{equation}\label{eq:potential_defn}
    \begin{aligned}
        \tilde{\Phi}(x_0, \gamma(x_0)) =& \sum_{i \in \Nc} \left[ L^{ii}_T(x^i_T) + \sum_{k=0}^{T-1} L^{ii}_k(x^i_k,u^i_k) \right] \\
        &+ \sum_{i, j \in \Nc, i < j} \sum_{H=0}^T \tilde{L}^{ij}_H(x^i_H, x^j_H).
    \end{aligned}
    \end{equation}
\end{thm}

We prove \cref{thm:sre_exact_PG} in Appendix~\ref{ssec:proofs_sre_dpg}.
Because the strategically robust costs $\tilde{J}^i$ form an exact dynamic potential game, every minimizer of the potential function \eqref{eq:potential_defn} is a strategically robust equilibrium.
Finally, two comments on the symmetry.
First, \cref{assume:monotonicity} implies $L^{ij}_H(x^i_H, x^j_H) = L^{ji}_H(x^j_H, x^i_H)$. Relaxing this assumption in certain ways (e.g., allowing agent-dependent weights)
would still yield a weighted dynamic potential game~\cite{bhatt2025strategic}.
Second, we also hypothesize that relaxing \cref{assume:deviation} such that $\epsilon^{i,H} \neq \epsilon^{j,H}$ under certain conditions could  yield an ordinal potential game. Ordinal and weighted potential games, with appropriate assumptions, have convergence guarantees~\cite{monderer_potential_1996}.
\begin{figure*}[!t]
    \centering
    \begin{subfigure}[b]{0.245\textwidth}
        \centering
        \includegraphics[width=\linewidth]{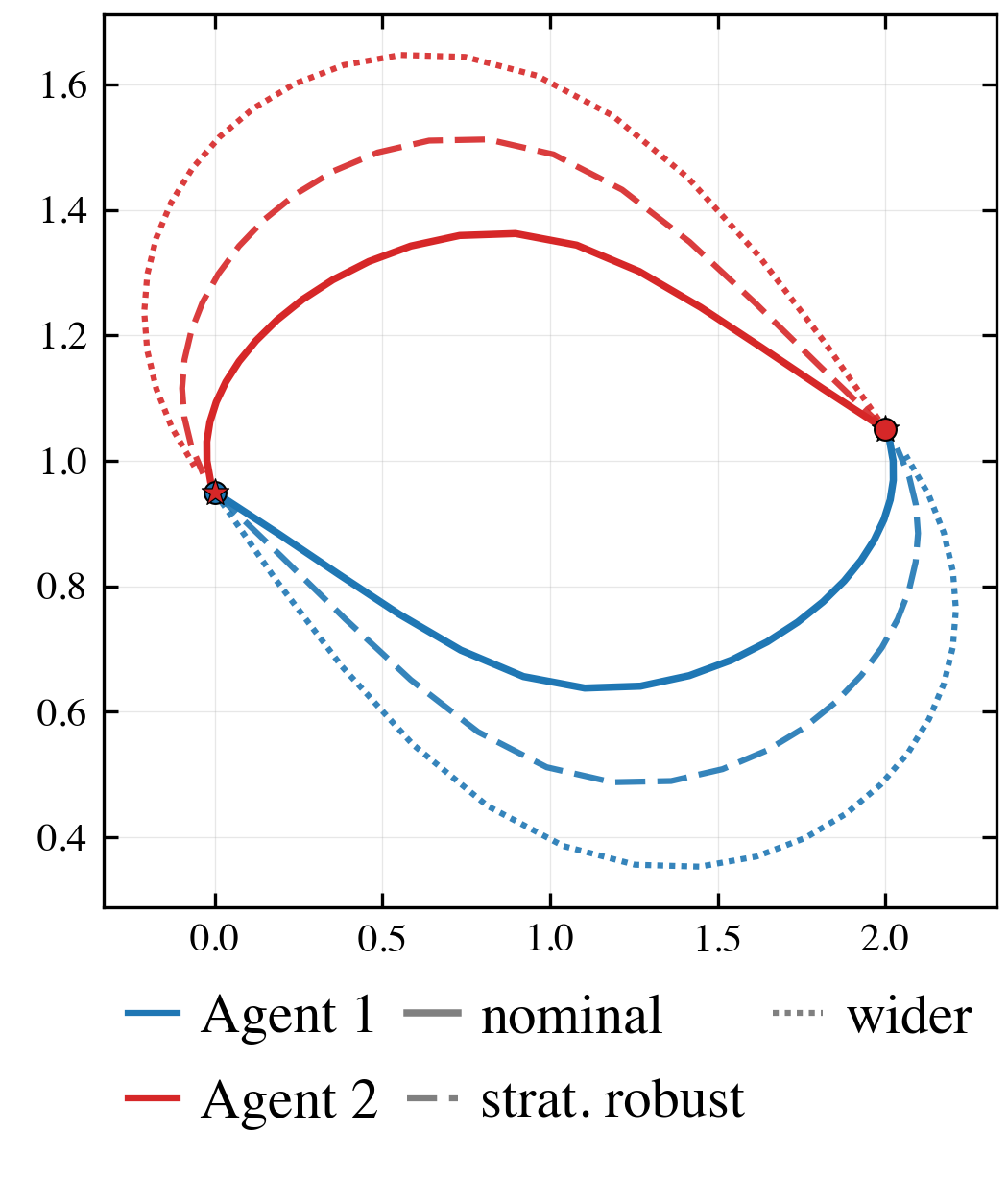}
        \caption{Head-on scenario.}
        \label{fig:headon}
    \end{subfigure}
    \hfill
    \begin{subfigure}[b]{0.245\textwidth}
        \centering
        \includegraphics[width=\linewidth]{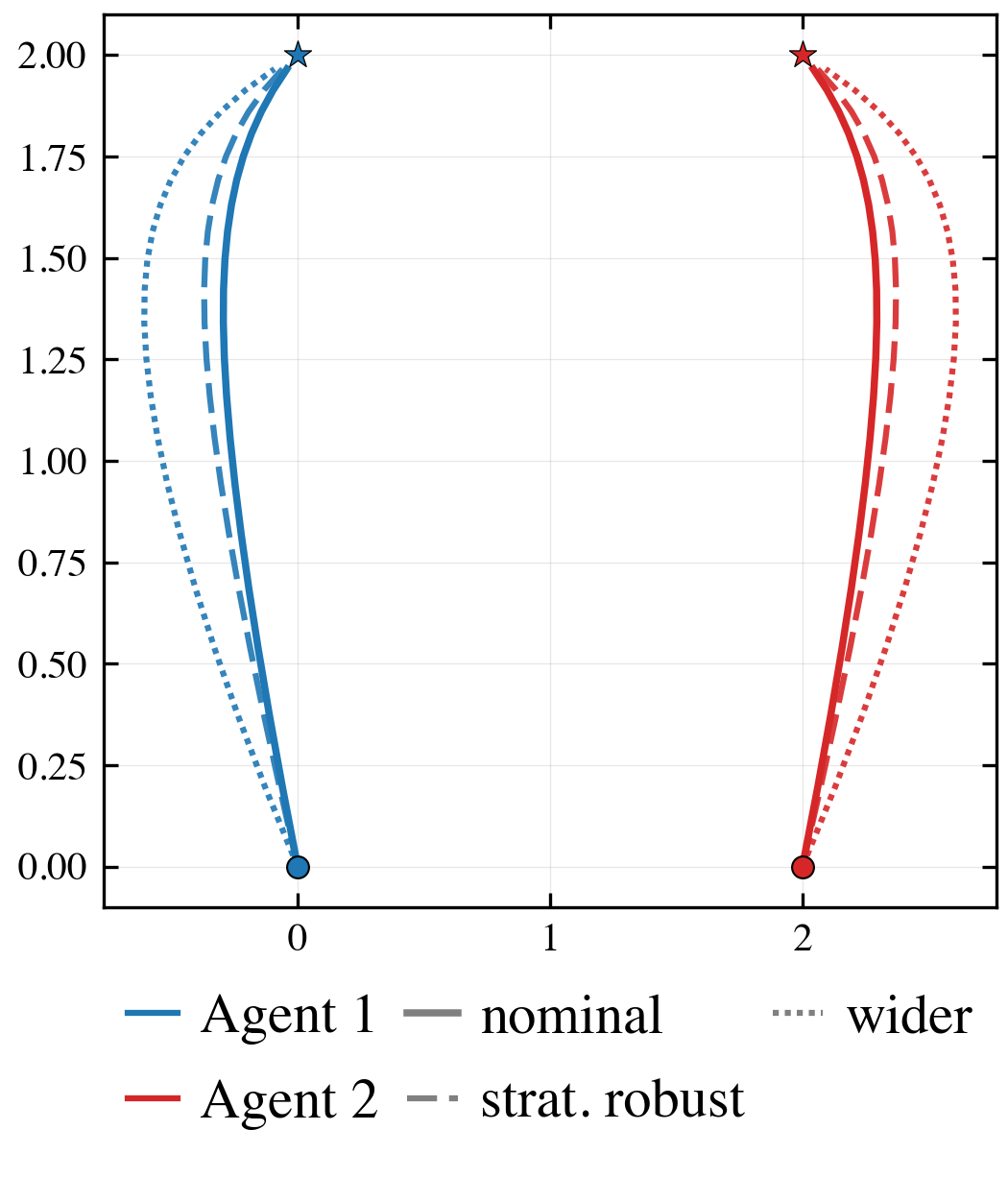}
        \caption{Parallel scenario.}
        \label{fig:parallel}
    \end{subfigure}
    \hfill
    \begin{subfigure}[b]{0.245\textwidth}
        \centering
        \includegraphics[trim=0cm 0cm 0cm 0cm, clip,width=\linewidth]{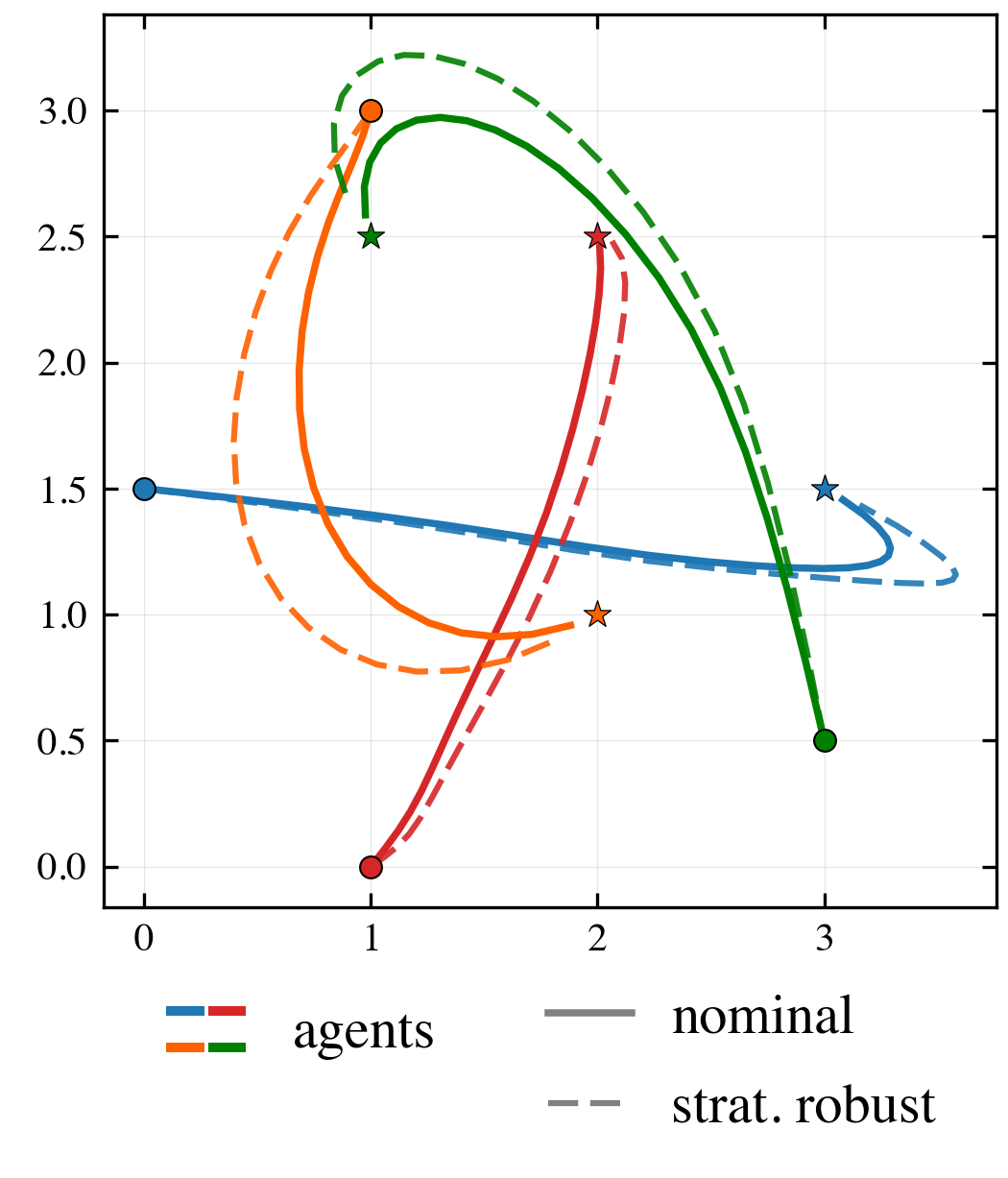}
        \caption{4-agent scenario.}
        \label{fig:4_multiagent}
    \end{subfigure}
    \begin{subfigure}[b]{0.245\textwidth}
        \centering
        \includegraphics[trim=0cm 0cm 0cm 0cm, clip, width=\linewidth]{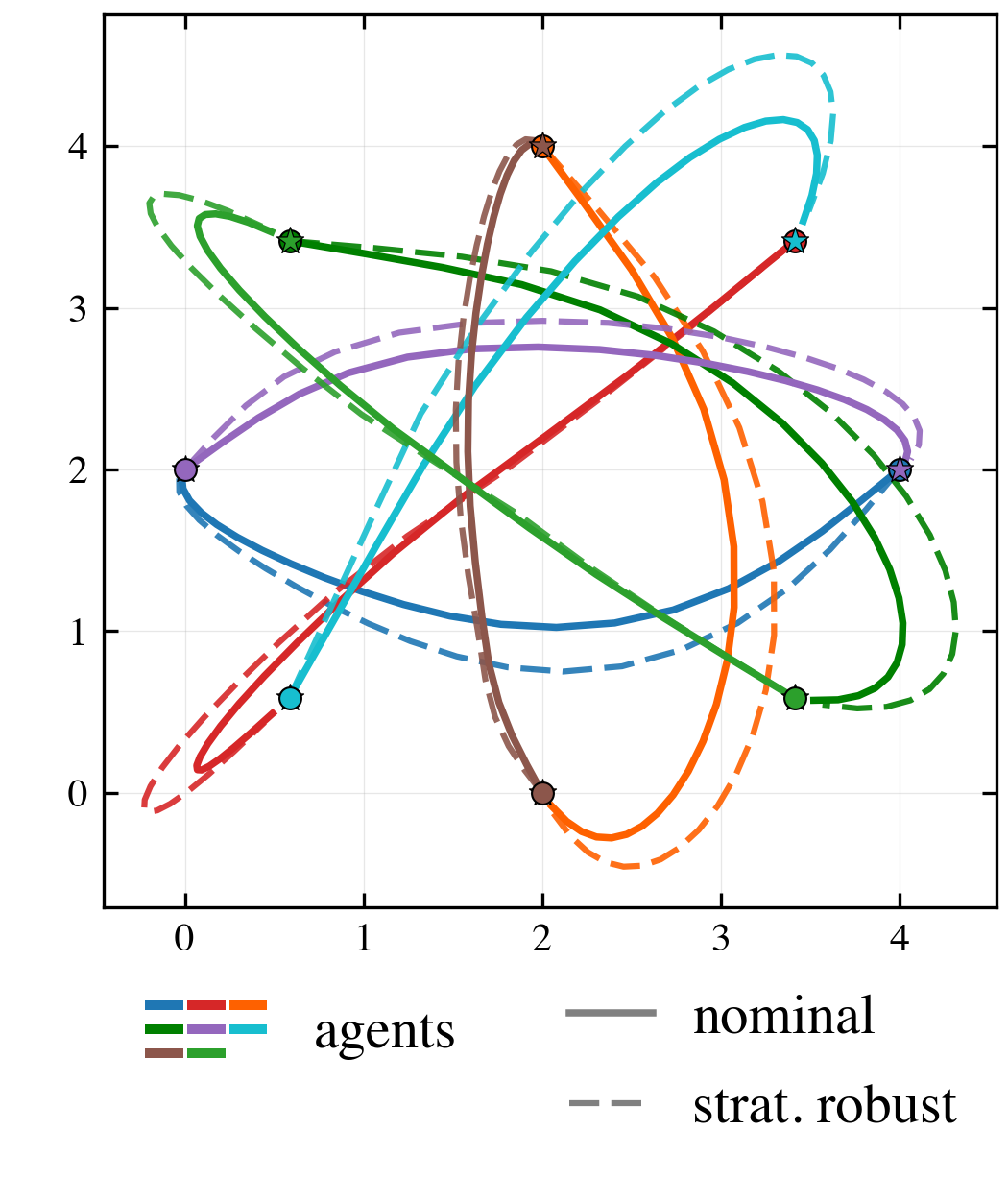}
        \caption{8-agent scenario.}
        \label{fig:8_multiagent}
    \end{subfigure}
    \caption{Comparison of trajectory optimization methods. Solid lines show nominal trajectories; different dashes show other methods. All trajectories are open-loop. (a)~Head-on: agents start at $(0,0.95)$ and $(2,1.05)$ and swap positions. (b)~Parallel: agents start at $(0,0)$ and $(2,0)$ and travel in the $y$-direction to $(0,2)$ and $(2,2)$. (c)~4 agents crossing a region. (d)~8 agents swapping antipodal positions on a circle.}
    \label{fig:all_scenarios}
\end{figure*}
\begin{table*}[!t]
    \centering
    \small
    \begin{minipage}[t]{0.36\textwidth}
        \centering
        \captionof{table}{Two-agent computation time}
        \label{tab:computation_time}
        \small
        \begin{tabular}{@{}l@{\hspace{0.25cm}}c@{\hspace{0.25cm}}c@{}}
        \toprule
        {\footnotesize $\times$nominal (s)} & \textbf{Head‑on} & \textbf{Parallel} \\
        \midrule
        \textbf{Nominal}     & $1.00\times$ (0.045s) & $1.00\times$ (0.012s)\\
        \textbf{Wider}       & $0.89\times$ (0.040s) & $1.17\times$ (0.014s) \\
        \textbf{Strat.\ Robust}& $1.41\times$ (0.064s)  & $2.01\times$ (0.024s) \\
        \bottomrule
        \end{tabular}
    \end{minipage}%
    \hfill%
    \begin{minipage}[t]{0.25\textwidth}
        \centering
        \captionof{table}{Deviation from nominal}
        \label{tab:deviation}
        \small
        \begin{tabular}{@{}l@{\hspace{0.2cm}}c@{\hspace{0.2cm}}c@{}}
        \toprule
        & \textbf{Head‑on} & \textbf{Parallel} \\
        \midrule
        \textbf{Nominal}     & $0.000$ & $0.000$ \\
        \textbf{Wider}       & $0.516$ & $0.449$ \\
        \textbf{Strat.\ Robust}& $0.252$ & $0.109$ \\
        \bottomrule
        \end{tabular}
    \end{minipage}%
    \hfill%
    \begin{minipage}[t]{0.37 \textwidth}
        \centering
        \captionof{table}{Multi‑agent computation time}
        \label{tab:multiagent}
        \small
        \begin{tabular}{@{}l@{\hspace{0.2cm}}c@{\hspace{0.2cm}}c@{}}
        \toprule
        {\footnotesize $\times$nominal (s)}   & \textbf{4 agents} & \textbf{8 agents} \\
        \midrule
        \textbf{Nominal} & $1.00\times$ (0.095s) & $1.00\times$ (1.356s) \\
        \textbf{Wider} & $1.04\times$ (0.099s) & $1.57\times$ (2.125s) \\
        \textbf{Strat.\ Robust} & $1.99\times$ (0.190s) &  $1.45\times$ (1.973s) \\
        \bottomrule
        \end{tabular}
    \end{minipage}%
    \vspace{-1em}
\end{table*}

\subsection{Fast computation of the strategically robust distance cost}\label{subsec:fast_computation}

While attractive, minimizing the potential still entails a computational challenge: the mere evaluation of $\tilde L^{ij}_H$ requires solving an optimization problem.
We now derive a quasi-closed-form and computationally efficient solution to this worst-case problem in three steps:

\begin{algorithm}[!t]
  \caption{Strategically Robust Potential Game}\label{alg:sr_solver}
  \begin{algorithmic}[1]
  \Require Game $G$, dynamics $(A,B)$, deviation bounds $\{\epsilon^{j,H}\}$
  \State Precompute eigendecompositions $Q_H, \Sigma_{\sigma_H}$
  \Repeat
      \For{pairs $i<j$, timesteps $H = 1,\dots,T$}
          \State Set $\lambda_H=0$ or find the positive root of \eqref{eq:newtons_eq} via Newton's method \textit{(Step 3)}
          \State Compute $\delta z^{j,H}_H$ in closed form \textit{(Step 2)}
      \EndFor
      \State Take minimization step on $\tilde{\Phi}(x_0, \gamma(x_0))$
  \Until{convergence}
  \end{algorithmic}
  \end{algorithm}

\emph{Step 1 (Monotonicity) }
By monotonicity of $\mu$, we can maximize the distance cost by minimizing the squared distance in \eqref{eq:adv_max_defn} for a given $j, H$. Define the relative position $z_H = x^i_H - x^j_H$, and the deviations $\delta z^{j,H}_k = x^j_k - \hat{x}^{j,H}_k$ and $\delta u^{j,H}_k = u^j_k - \hat{u}^{j,H}_k$.
Plugging into \eqref{eq:adv_max_defn} gives
\begin{equation*}
\begin{aligned}
    \max_{\hat{\gamma}^{j,H}}  -\mu (\Vert x^i_H - \hat{x}^{j,H}_H \Vert^2)
    =&\max_{\delta u^{j,H}_k} -\mu (\Vert z_H + \delta z^{j,H}_H \Vert^2) \\
    =& -\mu\left(\min_{\delta u^{j,H}_k} \Vert z_H + \delta z^{j,H}_H \Vert^2\right),
\end{aligned}
\end{equation*}
where we used the monotonicity of $\mu$. Thus, we can focus on solving the inner minimization.
The inner minimization satisfies Slater's condition for $\epsilon^{j, H} > 0$, so strong duality holds. For fixed dual multiplier $\lambda_H\geq 0$, we therefore write its Lagrangian relaxation, with $\delta z^{j,H}_{0}=0$:
\vspace{-0.5em}
\begin{equation}\label{eq:adv_monotonic_defn}
\begin{aligned}
    \min_{\delta u^{j,H}_k} \quad& \Vert z_H + \delta z^{j,H}_H\Vert^2
    + \lambda_H\left(\sum_{k = 0}^{H-1} \Vert \delta u^{j,H}_k \Vert^2 - {\epsilon^{j,H}}^2\right) \\
    \text{s.t.} \quad & \delta z^{j,H}_{k+1} = A \delta z^{j,H}_k + B \delta u^{j,H}_k \; \forall k \in [0, H-1] .
\end{aligned}
\end{equation}

\emph{Step 2 (Optimal Control) }
For fixed dual multiplier $\lambda_H\geq 0$, \eqref{eq:adv_monotonic_defn} is a quadratic optimization problem that gives us $\delta u^{j,H}_k$ and $\delta z^{j,H}_H$ in closed form, where $M_H = [A^{H-1}B, \dots, AB, B]$ (see~Appendix~\ref{subsec:fast_computation_details}):
\begin{equation*}
\begin{aligned}
    \delta z^{j,H}_H &=  M_H \delta u^{j, H} \\
    &= - M_H \trans{M_H} \left( M_H \trans{M_H} + \lambda_H I \right)^{\dagger} z_H .
\end{aligned}
\end{equation*}
Here ${}^{\dagger}$ denotes the Moore--Penrose pseudoinverse, which equals the ordinary inverse when $\lambda_H>0$. We then decompose $M_H \trans{M_H}$ into its eigendecomposition $Q_H \Sigma_{\sigma_H} \trans{Q_H}$, which can be precomputed for all $H$.

\emph{Step 3 (Updating $\lambda_H$) }
We can efficiently compute the dual multipliers $\lambda_H$ by first testing whether $\lambda_H=0$ is optimal; otherwise, we take the eigendecomposition of $M_H \trans{M_H}$ and solve the first-order condition of the dual using Newton iterations, as shown in~Appendix~\ref{subsec:fast_computation_details}. The final algorithm is summarized in \cref{alg:sr_solver}.

%% file: sections/4_results.tex
\section{Results}\label{sec:results}

We now illustrate strategic robustness on four multi-agent trajectory optimization scenarios, with two, four, and eight agents.
We compare three different methods, with $\eta = 10^{-8}$:
\begin{itemize}
    \item \textit{Nominal}:
    We compute the open-loop Nash equilibria by solving the exact dynamic potential game formed by $J^i$ in \eqref{eq:nominal_cost_function}, with $L^{ij}_k(x^i_k,x^j_k) = -2 \ln(\Vert x^i_k - x^j_k\Vert^2 + \eta)$.

    \item \textit{Strategically robust}:
    We compute the open-loop strategically robust equilibria by solving the exact dynamic potential game formed by $\tilde{J}^i$ using~\cref{alg:sr_solver}, with $L^{ij}_k(x^i_k,x^j_k) = -2 \ln(\Vert x^i_k - x^j_k \Vert^2 + \eta)$ and the robustness parameter $\epsilon^{j,H} = 2$ for all players and times.

    \item \textit{Wider}: We compute the open-loop Nash equilibria with increased distance cost $L^{ij}_k(x^i_k,x^j_k) = -5 \ln(\Vert x^i_k - x^j_k \Vert^2 + \eta)$. This adds robustness by increasing the distance between agents in all scenarios compared to the ``nominal'' method.
\end{itemize}
We used a direct single shooting method (cf. \cite{diehl2006fast}) to optimize the problem, solved using \texttt{scipy}'s SLSQP solver \cite{2020SciPy-NMeth}, which we provide with the function and its gradient (the Hessian is instead approximated numerically). Every example uses single integrator dynamics in $\R^2$, with $A = I, B = 0.1 I$, and $T = 20$; positions, times, and budgets are in normalized units. The private cost is $L^{ii}_k = \trans{(x^i_k - x^i_f)} Q_k(x^i_k - x^i_f) + \trans{u^i_k} R u^i_k$, with $Q_k = I$ for all $k\in\{0,\ldots,T-1\}$, $Q_T=150I$, and $R=I$.
The strategically robust method is initialized by first solving the nominal potential game; its reported computational time includes that initialization.\footnote{The code can be found at \url{https://github.com/victor-qin/strategically_robust_potential_trajopt}}

\subsection{Two-player system}

We consider the two scenarios in \cref{fig:headon,fig:parallel}, where two agents are traveling either head-on or in parallel.
To start, we observe that the strategically robust trajectories are wider compared to the nominal trajectories.
This is a direct consequence of strategic robustness: since agents protect against misbehavior by other agents, they commit to wider trajectories to reduce the collision risk.

\paragraph*{Increasing the collision parameter}
It is natural to ask if strategic robustness can be reproduced by a simple increase in the collision parameter, as in the ``wider'' method.
We argue here that strategic robustness offers protection in a targeted way.
Indeed, when comparing the ``strategically robust'' and ``wider'' methods, we observe that the effect of strategic robustness differs between \cref{fig:headon,fig:parallel}.
In~\cref{fig:headon}, the collision risk is concrete and strategic robustness leads to significantly wider trajectories.
In~\cref{fig:parallel}, strategic robustness instead leaves the trajectories nearly unchanged, as it reasons that a significantly larger deviation in control inputs is needed for a collision to occur.
%
Merely increasing the weight of the distance cost instead leads to approximately the same effect in both settings.
This qualitative observation can be made quantitative by inspecting the deviation of the trajectories from the nominal Nash equilibrium trajectory in~\cref{tab:deviation}, calculated as the trapezoidal approximation of the area between trajectories.
Strategic robustness deviates $2.3\times$ more from the nominal trajectory in the head-on scenario than in the parallel one, whereas merely increasing the collision parameter deviates by a comparable amount in both scenarios.

\paragraph*{Runtime}
We measure the runtime of different methods in \cref{tab:computation_time}, using 100 runs for each scenario, reported as [multiple of nominal time] (median time in seconds).
The strategically robust method only modestly increases the runtime compared to simply solving for the nominal trajectory.

\subsection{A multi-agent system}

Finally, we test our strategically robust trajectory planner for four and eight agents. We plot only the nominal and strategically robust methods for clarity.
The resulting trajectories are shown in~\cref{fig:4_multiagent,fig:8_multiagent}, and the computational times are listed in~\cref{tab:multiagent}. Strategic robustness takes between $1.41\times$ and $2.01\times$ the time of solving the nominal trajectory across two to eight agents, and the factor does not grow with the number of agents $N$.
%

%% file: sections/5_conclusion.tex
\section{Concluding Remarks}\label{sec:conclusion}

We believe our approach of reframing collision avoidance through strategic robustness~\cite{lanzetti_nicholas_strategically_2025} offers interesting directions for building fast trajectory optimization methods with collision avoidance guarantees.
We mention a few.
First, we have assumed that agents have identical dynamics, interagent costs, and robustness levels. As discussed in \cref{ssec:sre_as_pg}, we expect these assumptions can be relaxed to form weighted or ordinal dynamic potential games~\cite{monderer_potential_1996}.
Second, strategic robustness protects against the positions another agent could reach within its budget, pointing towards a separation certificate like those of backward reachability arguments. Subtracting a constant from the squared distance in our log-barrier cost (e.g., $-c\ln(\Vert x^i_k -x^j_k \Vert^2 - \Delta^2)$ for $\Delta>0$) would enforce a minimum separation, similar to control barrier functions.
Finally, we assumed linear dynamics and used \texttt{scipy}'s SLSQP. Integrating our method to work with dedicated solvers such as ALTRO~\cite{howell_altro_2019} (see~\cite{bhatt_efficient_2023, bhatt2025strategic}) or distributed potential iLQR~\cite{williams_distributed_2023} could facilitate extensions to nonlinear dynamics and constraints.

%% file: sections/a_proof.tex
\subsection{Proof of Theorem~\ref{thm:sre_exact_PG}}\label{ssec:proofs_sre_dpg}

It suffices to show that $\tilde{L}^{ij}_H(x^i_H, x^j_H) =  \tilde{L}^{ji}_H(x^j_H, x^i_H)$ for all $i < j \in \Nc, \, H \in [0, T]$, because then the terms of $\tilde{\Phi}$ from \eqref{eq:potential_defn} involving agent $i$ are exactly $\tilde{J}^i$. Define $\Theta^i \coloneqq \tilde{J}^i - \tilde{\Phi}$:
\begin{equation*}
\begin{aligned}
    \Theta^i(x^{-i}_0, \gamma^{-i}(x_0)) =& -\!\!\!\!\sum_{l \in \Nc \setminus \{i\}} \left[ L^{ll}_T(x^l_T) + \sum_{k=0}^{T-1} L^{ll}_k(x^l_k, u^l_k) \right] \\
    &- \!\!\!\! \sum_{p,l \in \Nc, p,l \neq i, p < l} \sum_{H=0}^{T} \tilde{L}^{pl}_H(x^p_H, x^l_H) .
\end{aligned}
\end{equation*}
Because dynamics are decoupled, $\Theta^i$ is a dummy function independent of $\gamma^{i}(x_0)$.
Thus $\{\tilde{J}^i\}_{i \in \Nc}$ forms an exact dynamic potential game by \cref{lem:coord_dummy}.
Next, we show $\tilde{L}^{ij}_H(x^i_H, x^j_H) =  \tilde{L}^{ji}_H(x^j_H, x^i_H)$ using \cref{assume:monotonicity,assume:deviation}.
Recall $\tilde{L}^{ij}_H(x^i_H, x^j_H)$ given by \eqref{eq:adv_max_defn}:
\begin{equation}\label{eq:new_adv_problem}
\begin{aligned}
    \max_{\hat{\gamma}^{j,H}} \quad & L^{ij}_H(x^i_H, \hat{x}^{j,H}_H) \\
    \text{s.t.} \quad & \hat{x}^{j,H}_{k+1} = A \hat{x}^{j, H}_k + B \hat{u}^{j, H}_k \; \forall k \in[0, H-1]
    \\
    & \hat{x}^{j,H}_{0}=x^j_0 \\
    & \sum_{k=0}^{H-1} \Vert u^j_k - \hat{u}^{j, H}_k \Vert^2 \leq {\epsilon^{j,H}}^2 .
\end{aligned}
\end{equation}
Define $\delta z^{j,H}_k = x^j_k - \hat{x}^{j,H}_k$ and $\delta u^{j,H}_k = u^j_k - \hat{u}^{j,H}_k$ to rewrite the dynamics as
\begin{equation*}
\begin{aligned}
    \hat{x}^{j,H}_{k+1} - x^j_{k+1} &= A (\hat{x}^{j,H}_k - x^j_k) + B (\hat{u}^{j,H}_k - u^j_k) \\
    \delta z^{j,H}_{k+1} &= A \delta z^{j,H}_k + B \delta u^{j,H}_k ,
\end{aligned}
\end{equation*}
with the initial condition $\delta z^{j,H}_{0}=0$ as $\hat{x}^{j,H}_0=x^j_0$.
Since $z_H = x^i_H - x^j_H$ is constant with respect to the adversarial optimization of \eqref{eq:new_adv_problem}, we can rewrite it as
\begin{equation}\label{eq:i_adv_problem}
\begin{aligned}
    \max_{\delta u^{j,H}_k} \quad &  -\mu(\Vert z_H + \delta z^{j,H}_H \Vert^2) \\
    \text{s.t.} \quad
    & \delta z^{j,H}_{k+1} = A \delta z^{j,H}_k + B \delta u^{j,H}_k \quad \forall k \in [0, H-1] \\
    & \delta z^{j,H}_0=0 \\
    & \sum_{k = 0}^{H-1} \Vert \delta u^{j,H}_k \Vert^2 \leq {\epsilon^{j,H}}^2.
\end{aligned}
\end{equation}
Second, we consider $\tilde{L}^{ji}_H(x^j_H, x^i_H)$, defined similarly to \eqref{eq:new_adv_problem}.
Following similar steps, define $\delta z^{i,H}_k = \hat{x}^{i,H}_k-x^i_k$ and $\delta u^{i,H}_k = \hat{u}^{i,H}_k-u^i_k$,
so that $x_H^j-\hat x_H^{i,H}=-z_H-\delta z_H^{i,H}$ (the sign flip ensures a match with the definition of $z_H$ above).
Then, we can rewrite $\tilde{L}^{ji}_H(x^j_H, x^i_H)$ as
\begin{equation}\label{eq:j_adv_problem}
\begin{aligned}
    \max_{\delta u^{i,H}_k} \quad&  -\mu (\Vert z_H + \delta z^{i,H}_H\Vert^2) \\
    \text{s.t.} \quad
    & \delta z^{i,H}_{k+1} = A \delta z^{i,H}_k + B \delta u^{i,H}_k \quad \forall k \in [0, H-1] \\
    & \delta z^{i,H}_0 = 0 \\
    & \sum_{k = 0}^{H-1} \Vert \delta u^{i,H}_k \Vert^2 \leq {\epsilon^{i,H}}^2.
\end{aligned}
\end{equation}
The case $H=0$ is trivial by $\hat{x}^{j,0}_0 = x^j_0$. Given that $\epsilon^{i,H} = \epsilon^{j,H}$ by~\cref{assume:deviation} and $(A, B)$ are identical by~\cref{assume:dynamics}, the problems \eqref{eq:i_adv_problem} and \eqref{eq:j_adv_problem} are the same, which implies that $\tilde{L}^{ij}_H(x^i_H, x^j_H) = \tilde{L}^{ji}_H(x^j_H, x^i_H)$ for all $H \in [0, T]$.

%% file: sections/a_derivation.tex

\allowdisplaybreaks[3]

\subsection{Details on Section \ref{subsec:fast_computation}} \label{subsec:fast_computation_details}

\emph{Step 2 (Optimal Control) } Consider \eqref{eq:adv_monotonic_defn} for timesteps $H \in [1, T]$, assuming $\epsilon^{j,H} > 0$ with fixed $\lambda_H \geq 0$ and $\delta z^{j,H}_0 = 0$.
Let $M_H = [A^{H-1}B, \dots, AB, B]$ and $\delta u^{j,H} $ be the vertical concatenation of $\delta u^{j,H}_0, \dots, \delta u^{j,H}_{H-1}$, so that $\delta z^{j,H}_H = M_H \delta u^{j, H}$.
We can rewrite \eqref{eq:adv_monotonic_defn} as
\begin{equation}\label{eq:adv_simplified_defn}
    \min_{\delta u^{j,H}}  \Vert z_H + M_H \delta u^{j, H} \Vert^2 + \lambda_H \left( \trans{\delta u^{j,H}} \delta u^{j,H} - {\epsilon^{j,H}}^2\right).
\end{equation}
For $\lambda_H > 0$, the objective is strictly convex, and setting its gradient to zero gives the unique input minimizer:
\begin{equation*}
\begin{aligned}
    \delta u^{j, H} &= - \left( \trans{M_H} M_H + \lambda_H I \right)^{-1} \trans{M_H} z_H \\
    &= - \trans{M_H} \left( M_H \trans{M_H} + \lambda_H I \right)^{-1} z_H .
\end{aligned}
\end{equation*}
Substituting into $\delta z^{j,H}_H$ gives
\begin{equation*}
    \delta z^{j,H}_H = - M_H \trans{M_H} \left( M_H \trans{M_H} + \lambda_H I \right)^{-1} z_H .
\end{equation*}
The matrix $M_H \trans{M_H}$ is the positive semidefinite controllability Gramian.
We can compute its eigendecomposition offline as $Q_{H} \Sigma_{\sigma_H}\trans{Q_{H}}$, where $\Sigma_{\sigma_H} = \text{diag}(\sigma_{1,H}, \dots, \sigma_{n,H})$, $\sigma_{l,H}\geq 0$, and $Q_H$ is an orthogonal matrix.
For $\lambda_H > 0$, we can then rewrite $\delta z^{j,H}_H$ as
\begin{equation*}
    \delta z^{j,H}_H = -Q_H \Sigma_{\frac{\sigma_H}{\lambda_H + \sigma_H}} \trans{Q_H} z_H ,
\end{equation*}
where $\Sigma_{\frac{\sigma_H}{\lambda_H + \sigma_H}}$ is quickly formed online.

If $\lambda_H = 0$, the problem reduces to least squares. We select the minimum-norm input $\delta u^{j,H}= -M_H^{\dagger} z_H$, giving $\delta z^{j,H}_H = -M_H M_H^{\dagger} z_H$, where $M_H^{\dagger}$ is the Moore--Penrose pseudoinverse. We can also derive this using the limit as $\lambda_H \to 0^+$ and following the singular value decomposition.

\vspace{0.1em}
\emph{Step 3 (Updating $\lambda_H$) } We maximize the dual function $q(\lambda_H)$ over $\lambda_H \geq 0$. Substituting $\delta u^{j,H}$ from Step 2 into the Lagrangian and simplifying gives, for $\lambda_H > 0$,
\begin{equation}\label{eq:adv_plugged_in}
    q(\lambda_H) = \trans{z_H} Q_H \Sigma_{\frac{\lambda_H}{\lambda_H + \sigma_H}} \trans{Q_H} z_H - \lambda_H {\epsilon^{j,H}}^2 .
\end{equation}
Let $w=\trans{Q_H}z_H$. As $\lambda_H \to 0^+$, the ratio $\lambda_H/(\lambda_H + \sigma_{l,H})$ becomes one for $\sigma_{l,H}=0$ and zero otherwise, so $q(0) = \sum_{l:\sigma_{l,H} = 0} w_l^2$, the squared residual in unreachable directions.
The dual function $q(\lambda_H)$ is concave on $\lambda_H \geq 0$. An interior maximizer satisfies
\begin{equation}\label{eq:newtons_eq}
\begin{aligned}
    \pdv{q}{\lambda_H} &= \trans{w} \Sigma_{\frac{\sigma_H}{(\lambda_H + \sigma_H)^2}} w - {\epsilon^{j,H}}^2 \\
    &= \sum_{l=1}^{n} \frac{{w_l}^2 \sigma_{l,H}}{(\lambda_H + \sigma_{l,H})^2} - {\epsilon^{j,H}}^2 = 0 .
\end{aligned}
\end{equation}
This unique positive root exists if and only if $\sum_{l:\sigma_{l,H}>0} w_l^2/\sigma_{l,H} > {\epsilon^{j,H}}^2$. Then we can compute the root using Newton's method, initialized at $\lambda_H > 0$ with $q'(\lambda_H) > 0$. Otherwise, $\lambda_H=0$ is optimal, and the adversary exactly intercepts the agent if and only if $z_H\in\operatorname{range}(M_H)$.